\documentclass[12pt,preprint]{aastex}
\usepackage{amssymb}
\newcommand{\rcsB}{RCS043938-2904.7}

\def\kms{km s$^{-1}$}
\def\bgc{B$_{gcR}$}

\begin{document}

\title{Evidence for line-of-sight structure in a comparison of X-ray and optical observations of the high-redshift cluster \rcsB}

\author{Benjamin Cain}
\affil{MIT Kavli Institute for Astrophysics and Space Research, Massachusetts Institute of Technology, 77 Massachusetts Ave., Cambridge, MA, 02139, USA}
\email{bcain@mit.edu}

\author{David G. Gilbank}
\affil{Department of Astronomy and Astrophysics, University of Toronto, 50 St. George St., Toronto, Ontario, M5S 3H4, Canada}
\email{gilbank@astro.utoronto.ca}

\author{M.W. Bautz}
\affil{MIT Kavli Institute for Astrophysics and Space Research, Massachusetts Institute of Technology, 77 Massachusetts Ave., Cambridge, MA, 02139, USA}
\email{mwb@space.mit.edu}

\author{A. Hicks}
\affil{Department of Astronomy, University of Virginia, P.O. Box 400325 Charlottesville, VA, 22904, USA}
\email{ahicks@alum.mit.edu}

\author{H. K. C. Yee}
\affil{Department of Astronomy and Astrophysics, University of Toronto, 50 St. George St., Toronto, Ontario, M5S 3H4, Canada}
\email{hyee@astro.utoronto.ca}

\author{M. Gladders}
\affil{Department of Astronomy and Astrophysics, University of Chicago, 5640 S. Ellis Ave., Chicago, IL, 60637, USA}
\email{gladders@oddjob.uchicago.edu}

\author{E. Ellingson}
\affil{Center for Astrophysics and Space Astronomy, University of Colorado at Boulder, Campus Box 389, Boulder, CO, 80309, USA}
\email{Erica.Ellingson@colorado.edu}

\author{L.F. Barrientos}
\affil{Departamento de Astronom\'{\i}a y Astrof\'{\i}sica, Universidad Cat\'{o}lica de Chile, Avenida Vicu\~{n}a Mackenna 4860, Casilla 306, Santiago 22, Chile}
\email{barrientos@astro.puc.cl}

\and

\author{G.P. Garmire}
\affil{Department of Astronomy and Astrophysics, 525 Davey Lab, Pennsylvania State University, University Park, PA, 16802, USA}
\email{garmire@astro.psu.edu}
\shortauthors{Cain et al.}

\begin{abstract}
We present new \emph{Chandra} observations of a high redshift ($z$$\sim$1) galaxy cluster discovered in the Red-Sequence Cluster Survey (RCS): \rcsB.  X-ray luminosity measurements and mass estimates are consistent with $L_X$--$T_X$ and $M_\delta$--$T_X$ relationships obtained from low-redshift data.  Assuming a single cluster, X-ray mass estimates are a factor of $\sim$10--100 below the red-sequence optical richness mass estimate.  Optical spectroscopy reveals that this cluster comprises two components which are close enough to perhaps be physically associated.  We present simple modeling of this two-component system which then yields an X-ray mass and optical richness consistent with expectations from statistical samples of lower redshift clusters.  An unexpectedly high gas mass fraction is measured assuming a single cluster, which independently supports this interpretation.  Additional observations will be necessary to confirm the excess gas mass fraction and to constrain the mass distribution.
\end{abstract}

\section{Introduction}
There is considerable interest in measuring the evolution of the galaxy cluster mass function with redshift.  In principle, such a measurement can test structure formation and constrain the cosmic expansion history.  One of the observational challenges in such an undertaking is to identify all clusters in a particular volume; the other is to reliably infer cluster mass from observable cluster properties, over a range of cluster redshifts.  Recent optical surveys, such as the Red-Sequence Cluster Survey \citep[RCS,][]{GY05}; and the Sloan Digital Sky Survey \citep[e.g.,][]{Kea07} have found large numbers of clusters at moderate to high redshift, with $0.2\lesssim z\lesssim 1$ and $0< z \lesssim 0.3$, respectively.  At the high redshift extent, where spectroscopic information can be difficult to obtain on a survey scale, various optical properties, such as richness or optical luminosity, can be taken to be mass proxies, and have been shown to be strongly correlated with observables understood to be related to cluster mass, such as galaxy velocity dispersion and X-ray temperature \citep[e.g.,][]{YE03, Lea03, Pea04, Lea06}.   In order to use optical richness (or any other optical property) as a mass proxy, understanding the evolution of the mass-richness relation is critical to extracting cosmological information, since any evolution in the relation must be separated from the evolution of the mass function.  In particular, some care must be taken when attempting to extend the mass/observable relation to high redshift as many of these relations are computed using local data.  

Previous work has shown that optically-selected, high-redshift clusters at a fixed optical richness are typically lower in temperature and underluminous in X-rays when compared to local clusters \citep{Hea05,Gea04,Lea02,Dea01}.  Since these clusters generally follow the local luminosity-temperature relationship, the low temperatures and luminosities are interpreted as an indication that the cluster galaxies are not in virial equilibrium with the X-ray emitting gas.  However, it is important to account for the significant scatter in the relation between optical richness and X-ray luminosity, as an Eddington-like bias will scatter points of lower than expected X-ray luminosity into an optically-selected sample \citep[e.g.,][]{Bea94}.  Observing these high redshift clusters is thus an avenue to study how clusters approach equilibrium.

\rcsB\ is part of a larger sample of high redshift, high richness RCS clusters selected for X-ray follow up observations.  X-ray properties of the sample as a whole are described in a forthcoming paper by \citet{Hea07}.  Here we present \emph{Chandra} X-ray observations and optical spectroscopy of this apparently rich, X-ray underluminous object.  In \S 2 we describe the X-ray data reduction and analysis.  In \S 3 we compare mass estimates from the X-ray data to the known mass-temperature and mass-richness relations in literature.  In \S 4 we describe the optical spectroscopy, in \S 5 we discuss our results and we summarize in \S 6.  Throughout the paper we take a standard $\Lambda$CDM cosmology with $H_0 = 70$ km s$^{-1}$ Mpc$^{-1}$, $\Omega_m = 0.3$ and $\Omega_\Lambda = 0.7$.  At the cluster redshift, $z=0.9558$, 1\arcsec\ corresponds to a metric distance of 6.64 kpc.  Errors are indicated with 90\% confidence intervals unless otherwise noted.

\section{X-ray observations \& data analysis}

\subsection{Reduction}

The X-ray data consists of two \emph{Chandra} ACIS-S3 observations.  A log of the observations of \rcsB\ is presented in Table \ref{obstable}.  Both observations were performed in the VFAINT mode, and the CTI correction was applied with a focal plane temperature of -120$^{\circ}$C.  Data reduction was done using CIAO version 3.2 tools and CALDB 3.3.0 starting with level 2 event lists from the standard \emph{Chandra} pipeline.

The first observation of \rcsB\ (obsid 3577) had a highly variable background and the second observation was kindly provided by the Director of the \emph{Chandra} X-ray Center to compensate.  This second observation (obsid 4438) had a low background and was analyzed by standard lightcurve methods to select the acceptable background rate intervals, keeping 28.7 ks of the total 28.8 ks.  More care had to be taken with obsid 3577 because standard lightcurve methods eliminated more than three-quarters of the data as unusable.  

In order to maximize the useful portion of the data, we took the following approach.  We binned the data into 100-second segments and sorted them from lowest number of counts to highest number of counts.  We assume that the cluster flux is constant in time. This means that the signal-to-noise ratio is highest in the bin with the least number of counts and lowest in the bin with the most counts.  Suppose then that there are $S$ signal counts in each bin and $C_n$ total counts in the first $n$ bins.  Since $C_n$ monotonically increases with $n$, we want to find the lowest value of $n$ for which 
\begin{equation}
\frac{Sn}{\sqrt{C_n}}>\frac{S(n+1)}{\sqrt{C_{n+1}}},
\end{equation}
or equivalently,
\begin{equation}
n>\frac{1}{\sqrt{\frac{C_{n+1}}{C_n}}-1},
\end{equation}
meaning that adding the data in the $(n+1)^{\textrm{th}}$ bin will reduce the total signal-to-noise ratio in our filtered data set.  This determined the time intervals we included in our final data set independent of the source count rate.  In total, 64.6 ks of the total 76.2 ks of data (85\%) from obsid 3577 were kept.  These data were combined with the second observation (obsid 4438) using the CIAO script \texttt{merge\_all} for a total good-data exposure time of 93.3 ks.  The same script was used to create an exposure map and exposure-corrected image.

The X-ray data were then filtered to the energy range of 0.3-7 keV and point sources were identified and removed using {\texttt wavdetect}.  Pulse-height spectra and associated response files were created from a 100 pixel (49.2$\arcsec$) radius circular region around the X-ray centroid.  Background spectra were obtained from another 100 pixel radius circular region over 400 pixels (196.8$\arcsec$) away from the cluster center, though still from the same merged image.  Care was taken to choose the background region in a location where the two observations overlapped.  Spectral analysis was performed using XSPEC version 12.2.0 \citep{A96}, and cluster emission was modeled as an optically-thin thermal plasma \citep{MGvdO85,MLvdO86}.  The redshift was fixed at the value determined from optical spectroscopy and the heavy element abundance was fixed at 0.3 times Solar abundance \citep{AG89}.  An attempt was made to determine the metal abundances, but the data do not provide a significant constraint.  The spectral model also included Galactic photoabsorption.  The nH FTOOL from NASA's HEASARC \citep{DL90} was used to fix the column density of HI.  Spectra were binned so that a minimum of 20 counts were in each energy bin.

\subsection{Temperature, surface brightness, mass profile \& gas mass fractions}

The best fit temperature, fluxes and luminosities (with 90\% confidence intervals) are listed in Table \ref{spectable}.  The bolometric luminosity was estimated by integrating the model spectra between 0.05 and 50 keV, and over the 49.2\arcsec\ radius aperture.  The corresponding metric aperture radius at the cluster redshift is 391 kpc for \rcsB.

We extracted a radial profile of the cluster emission from the exposure-corrected image.  Point sources were removed using the CIAO script \texttt{dmfilth} and a $\beta$-model \citep{CFF76} was fit to the data.  Data for the fit was taken from a circular aperture with a radius of 98.4\arcsec.  This aperture was split into circular annuli with 4.92\arcsec\ thicknesses and fit using the Sherpa program included in the CIAO software  The $\beta$-model for the surface brightness is:
\begin{equation}
I(r)=I_0 \left(1+ \left(\frac{r}{r_0}\right)^2\right)^{-3 \beta + \frac{1}{2}}.
\end{equation}
We used the observed flux within the 49.2\arcsec\ radius spectral aperture to express the normalization in terms of physical units.  We also determine the central electron density, $n_{e0}$, using the thermal-plasma emission model discussed above and assuming that the plasma is isothermal throughout the intra-cluster medium (ICM).  The $\beta$-model fitting results and 90\% confidence intervals are presented in Table \ref{betatable}.

Assuming that the cluster is isothermal and in hydrostatic equilibrium, we can infer the mass distribution \citep[e.g.,][]{N05} of 
\begin{equation}
M(r)=\frac{3 \beta T_X}{G \mu m_p}\frac{r^3}{r^2+r_0^2},
\end{equation}
where $\mu$ is taken at 0.59 for fully ionized hydrogen and helium of cosmic abundances.  To test the validity of the isothermality assumption we fit temperatures for two portions of the spectral aperture: A circular region about the center of the emission and an annulus immediately outside the circle. The border between these sub-regions was chosen so that the annulus and the circle each contain roughly the same number of source counts.  Results are consistent with the isothermal assumption but provide no additional constraint.

We extrapolated the mass profile to obtain $r_{\delta}$, the radius at which the spherically-averaged cluster density is $\delta$ times the critical density at the cluster redshift, and $M_{\delta}$, the cluster mass within $r_{\delta}$.  We evaluate these parameters for $\delta = $200, 500 and 2500.  We detect cluster emission above background to a radius of approximately $r_{500}$, meaning that we see cluster emission from the majority of the area inside the aperture used for obtaining the spectral data ($r_{500}=38.9\arcsec$ versus the spectral aperture of 49.2$\arcsec$).  Values for $M_{\delta}$ and $r_{\delta}$ are listed in Table \ref{masstable}.  We find that though the X-ray temperature, and by extension the inferred mass, is low compared to other RCS clusters of similar richness, the measured core radius is comparable \citep{Hea07}.  Lastly, we estimate the cluster gas mass and gas mass fraction at these three density contrasts by applying our $\beta$-model for the gas density and comparing to our mass estimates.  Results from this analysis are presented in alongside the mass estimates in Table \ref{masstable}.

\section{Relations between observables}

\subsection{Mass-temperature relation}

\begin{figure}[h]
\plottwo{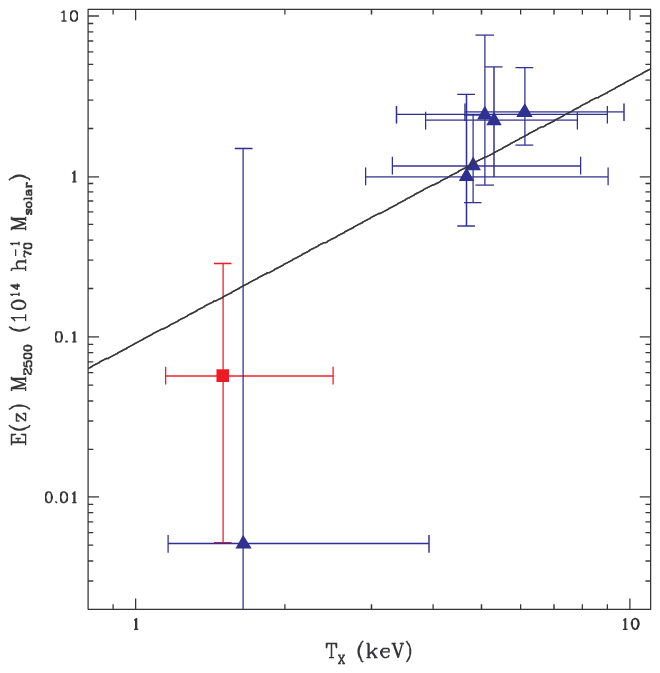}{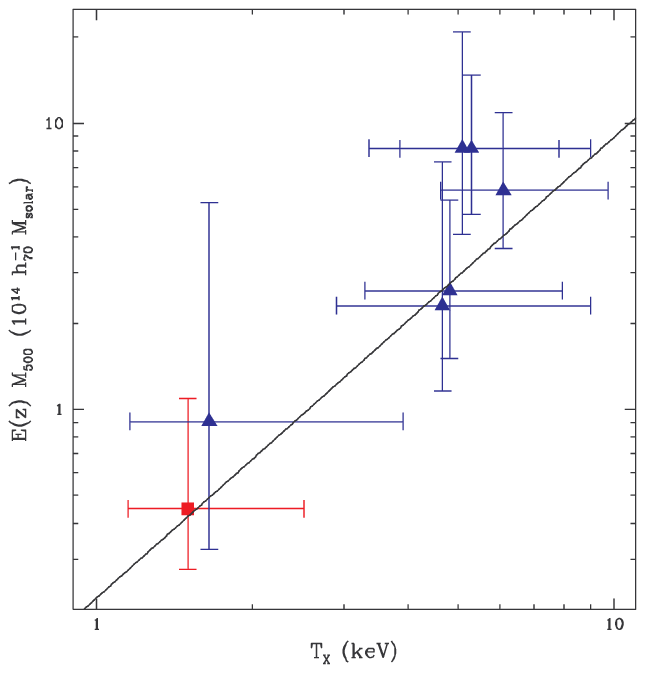}
\caption{Mass-Temperature Relation for $M_{2500}$ (at left) and $M_{500}$ (at right).  The solid lines are the relation obtained by \citet{Vea06}.  The red square point on each plot indicates \rcsB.  The blue triangle points indicate a selection of other RCS clusters which were similarly analyzed and are provided only for comparison.  A detailed analysis of the set of RCS clusters and their redshift evolution is beyond the scope of this paper and will be provided in \citet{Hea07}.}
\label{MTplot}
\end{figure}

We compare the values of $M_{\delta}$ calculated from the $\beta$-model to the mass-temperature relations for $M_{500}$ and $M_{2500}$ determined by \citet{Vea06} in Figure \ref{MTplot}.  This figure also includes data from six other RCS clusters, similarly analyzed, for comparison.  A full treatment of this larger sample will be included in \citet{Hea07}.  We scaled masses by $E(z) \equiv H(z)/H_0$ to account for redshift evolution expected in the self-similar model \citep[e.g.,][]{BN98}.  The $M$--$T_X$ relation is then parametrized as
\begin{equation}
E(z) M_\delta=M_0\ \left(\frac{T_X}{5 \ \textrm{keV}} \right)^\alpha \ 10^{14} \ h_{70}^{-1} \textrm{M}_{\odot},
\end{equation}
with $M_0$ and $\alpha$ taken directly from \citet{Vea06}:  For $\delta=500$, $M_0 = 2.93\pm0.16$ and $\alpha = 1.61\pm0.11$; for $\delta=2500$, $M_0 = 1.28\pm0.05$ and $\alpha = 1.64\pm0.06$.  \rcsB\ agrees with the relation (at 90\% confidence) for both $\delta = 500$ and $\delta = 2500$.

An additional mass estimator based on X-ray data which is known to have a tight correlation with cluster mass is the integrated pressure $Y_\delta\equiv M_{g,\delta}T_X$ \citep{Kea06,Nea07}.  We compare our cluster data to the relation obtained through simulations done by \citet{Kea06} and with the relation obtained from \emph{Chandra} data by \citet{Nea07}.  Our data agree  at the 90\% confidence level with both relations.

\subsection{Luminosity-temperature relation}

We adopt the luminosity-temperature parametrization of a simple power law and redshift evolution following the self-similar evolution model
\begin{equation}
E(z)^{-1}L_{X}=L_0 \ \left( \frac{T_X}{6\ \textrm{keV}}\right)^\alpha \ 10^{44} \ h_{70}^{-2}\  \textrm{erg s}^{-1}.
\end{equation}

\begin{figure}[h]
\plotone{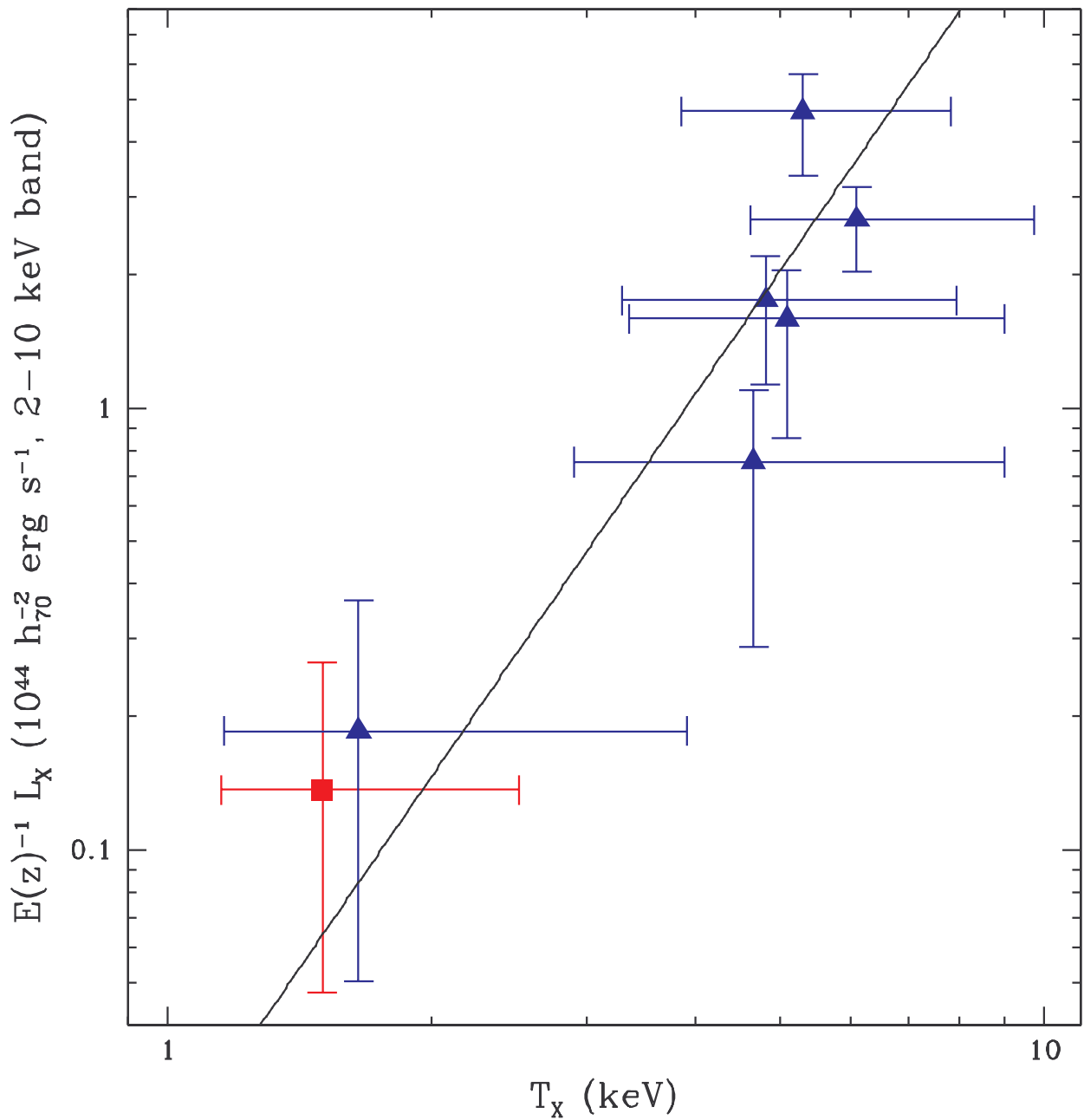}
\caption{Luminosity-Temperature Relation.  Luminosity is in the 2-10 keV band, extrapolated to the radius $r_{200}$.  Redshift evolution is scaled out according to the self-similar model.  The red square point indicates \rcsB.  The blue triangle points indicate a selection of other RCS clusters which were similarly analyzed and are provided only for comparison.  A detailed analysis of the set of RCS clusters and their redshift evolution is beyond the scope of this paper and will be provided in \citet{Hea07}.}
\label{LTplot}
\end{figure}

For the temperature dependence exponent and normalization, we take the local relation determined by \citet{AE99} for luminosity in the 2-10 keV band extrapolated to $r_{200}$.  This gives $\log_{10} L_0 = 0.54 \pm 0.03$ and $\alpha = 2.88 \pm 0.15$.  \citet{Mea06}, who analyzed 11 high-$z$ clusters observed with XMM and/or \emph{Chandra}, find this relation to be an acceptable fit to their data as well.  Their assumed redshift evolution includes an additional factor to account for evolution of the density contrast needed for virialization, but they note that it is dominated by the self-similar evolution factor.  Given the size of the errors in our measurements, we neglect this smaller term.   We compare our extrapolated luminosity in the 2-10 keV band with the relation in Figure \ref{LTplot}, which shows that the present data are consistent with the local relation.  Six other RCS clusters, whose complete treatment will be presented in \citet{Hea07}, are included in the figure for comparison.

\subsection{Mass-richness relation}

\begin{figure}[h]
\plotone{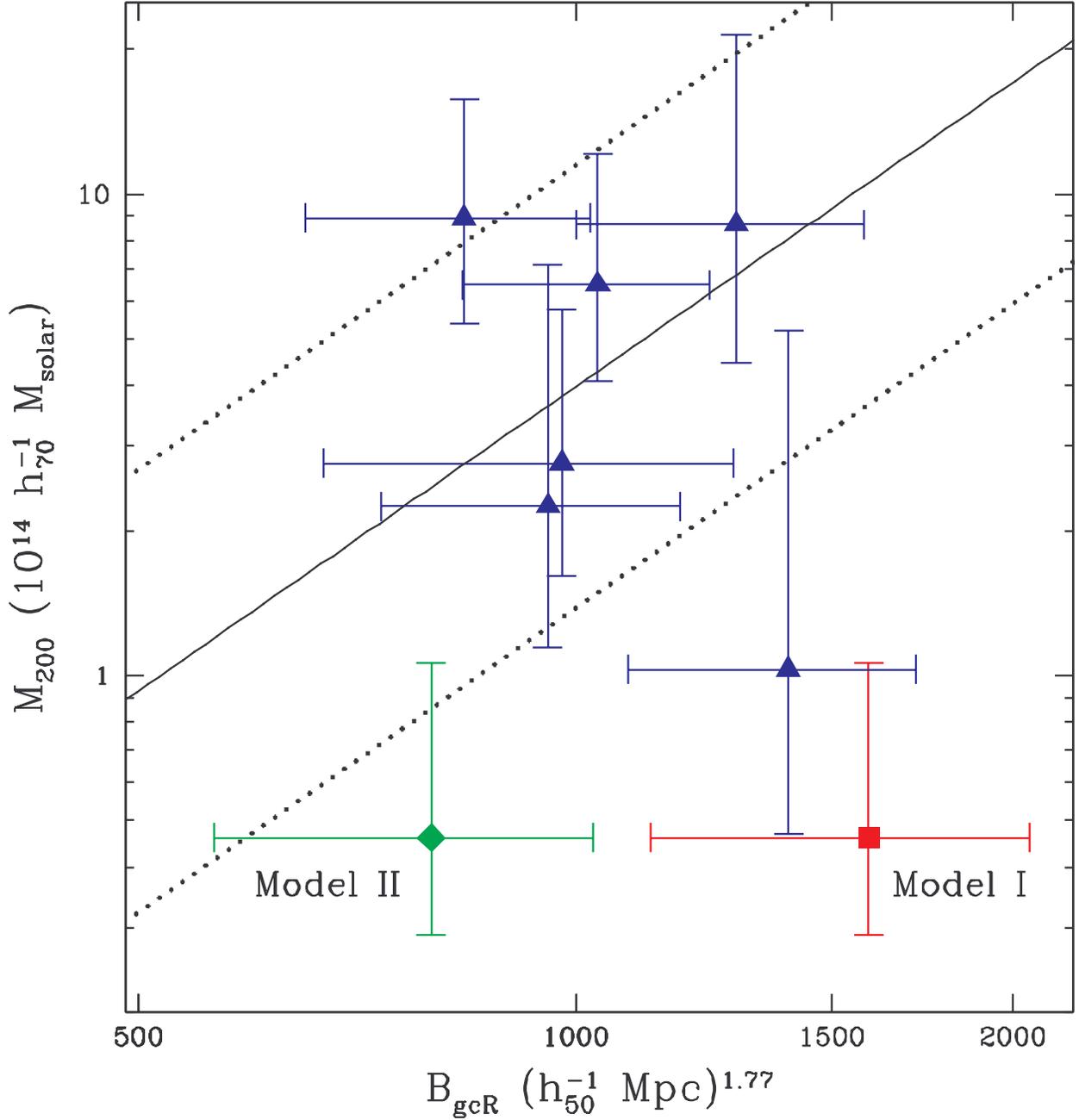}
\caption{The $M_{200}$--$B_{gcR}$ relation, plotting X-ray mass against richness. The solid line is the Blindert et al. relation and the dotted lines above and below are the $1\sigma$ scatter ($\pm0.46$ dex).  The red square point indicates \rcsB\ assuming that all the richness is associated with a single component (Model I).  The green diamond point indicates each of the two equal-mass components within \rcsB\ assuming Model II, with half the observed richness associated with each component.  The blue triangle points indicate a selection of other RCS clusters which were similarly analyzed and are provided only for comparison.  A detailed analysis of the set of RCS clusters and their redshift evolution is beyond the scope of this paper and will be provided in \citet{Hea07}.}
\label{MBfig}
\end{figure}

Optical richness can be quantified using the $B_{gc}$ parameter introduced by \citet{LS79}, which has been shown to correlate well with cluster velocity dispersion, as well as other observables which have been used for cluster mass proxies \citep{YE03}.  See \citet{YLC99} for a detailed review of richness measures.  The RCS measures a modified version of this parameter, $B_{gcR}$, considering only the red-sequence galaxies in their richness calculation as an attempt to better estimate the cluster mass \citep{GY05}.  The mass-richness relation is parametrized as
\begin{equation}\label{MBgceq}
M_{200} = 10^{A} B_{gcR}^\alpha (1+z)^\gamma\ 10^{14}\ \textrm{M}_{\odot},
\end{equation}
as in \citet{Gladders_ea07}.  The measured $B_{gcR}$ value for \rcsB\ of 1591$\pm494$ $(h_{50}^{-1}$ Mpc$)^{1.77}$ (also listed in Table \ref{betatable}) was derived from the RCS data using the method described in \citet{GY05}.  We use the relation obtained by \citet{Bea08} from a dynamical study of 33 low- to moderate-redshift ($z<$0.6) RCS clusters.  They found values of $A=-5.7\pm3.4$ and $\alpha=2.1\pm1.2$ (68\% confidence) and a $\pm 0.46$ dex scatter in mass at a constant richness.  This is consistent with the values measured by \citet{YE03} for the X-ray selected CNOC1 sample, over a similar redshift range.  The Blindert sample was limited in redshift and places no constraint on redshift evolution, so we take $\gamma \equiv 0$.  This is also consistent with the values and evolution inferred from the cosmological study of \citet{Gladders_ea07}.  In Figure \ref{MBfig} we compare our cluster to the Blindert relation.  Assuming that the spherically symmetric $\beta$-model is a good representation of the system, \rcsB\ is a significant outlier, with far too large a richness mass for its X-ray mass.  Optical spectroscopy measurements of \rcsB\ show evidence of significant substructure, as will be described in the next section.

\section{Optical spectroscopy}

Optical spectra have been obtained for galaxies in \rcsB\ \citep{Gilbank_ea07,Bea04}.  The details of the observations are given in these papers.  Briefly, the datasets comprise 34 spectroscopically confirmed members of the \rcsB\ system, with redshifts from LDSS-2 and IMACS on the 6.5-m Magellan Baade telescope and FORS2 on the VLT.  The spectra of \citet{Gilbank_ea07} were classified on a scale of 1-5 in terms of redshift quality.  For this discussion we take classes 1-3 to be secure redshifts and class 4 to be lower confidence redshifts.  Redshift measurements were not possible for class 5 spectra.  The VLT spectra of Barrientos et al.\ (2004) have been reclassified onto the same system.  To this sample we add recent spectroscopy from FORS2 on the VLT, similar to those described in \citet{Bea04}.  These data will be presented in Barrientos et al., in preparation.

\begin{figure}[h]
\plotone{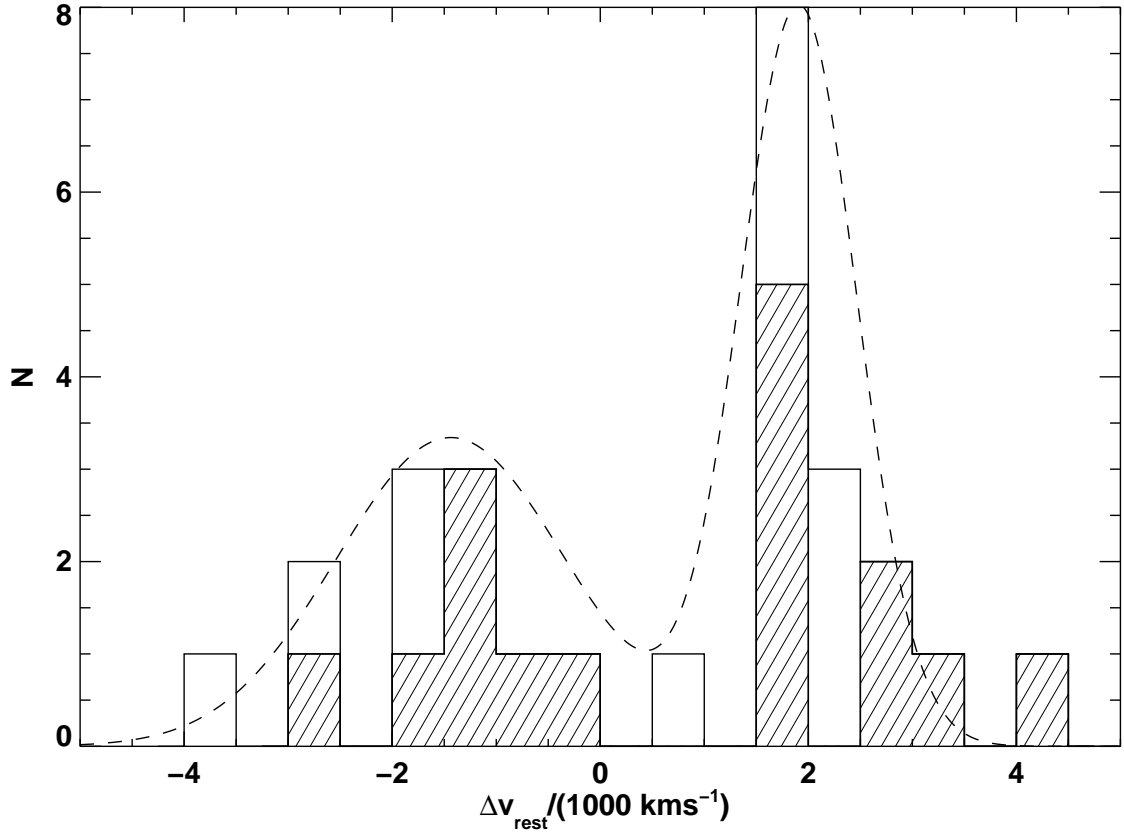}
\caption{Rest-frame velocity histogram centered around the mean velocity of the \rcsB, z$=$0.9558.  Hatched histogram shows secure redshifts and open histogram shows lower confidence measurements. The overplotted curves show best double-gaussian fits to these, solid and dashed curves respectively.  Uncertainties are sufficiently large that velocity dispersion measurements for each of the two components are consistent with equal velocity dispersions.}
\label{veldisfig}
\end{figure}

As discussed in \citet{Gilbank_ea07}, \rcsB\ displays a clear double peak in the redshift histogram at $z\sim0.96$.  Figure \ref{veldisfig} shows a redshift histogram centered on the mean redshift of the two components, $z=0.9558$.  Only galaxies within a radius of 5 arcminutes, corresponding to a projected distance of $\sim2.0\ h_{70}^{-1}$ Mpc, are considered.  Fitting a double-gaussian to the peaks yields central locations of the two peaks which are insensitive to the redshift quality class under consideration.  The separation of the two peaks in the mean rest frame ranges from 3000 \kms\ to 3250 \kms\ depending on the redshift quality chosen.  The difference between the two cases is comparable with the redshift uncertainty on a single measurement \citep[$\sim200$ \kms,][]{Gilbank_ea07}.

\begin{figure}[h]
\plotone{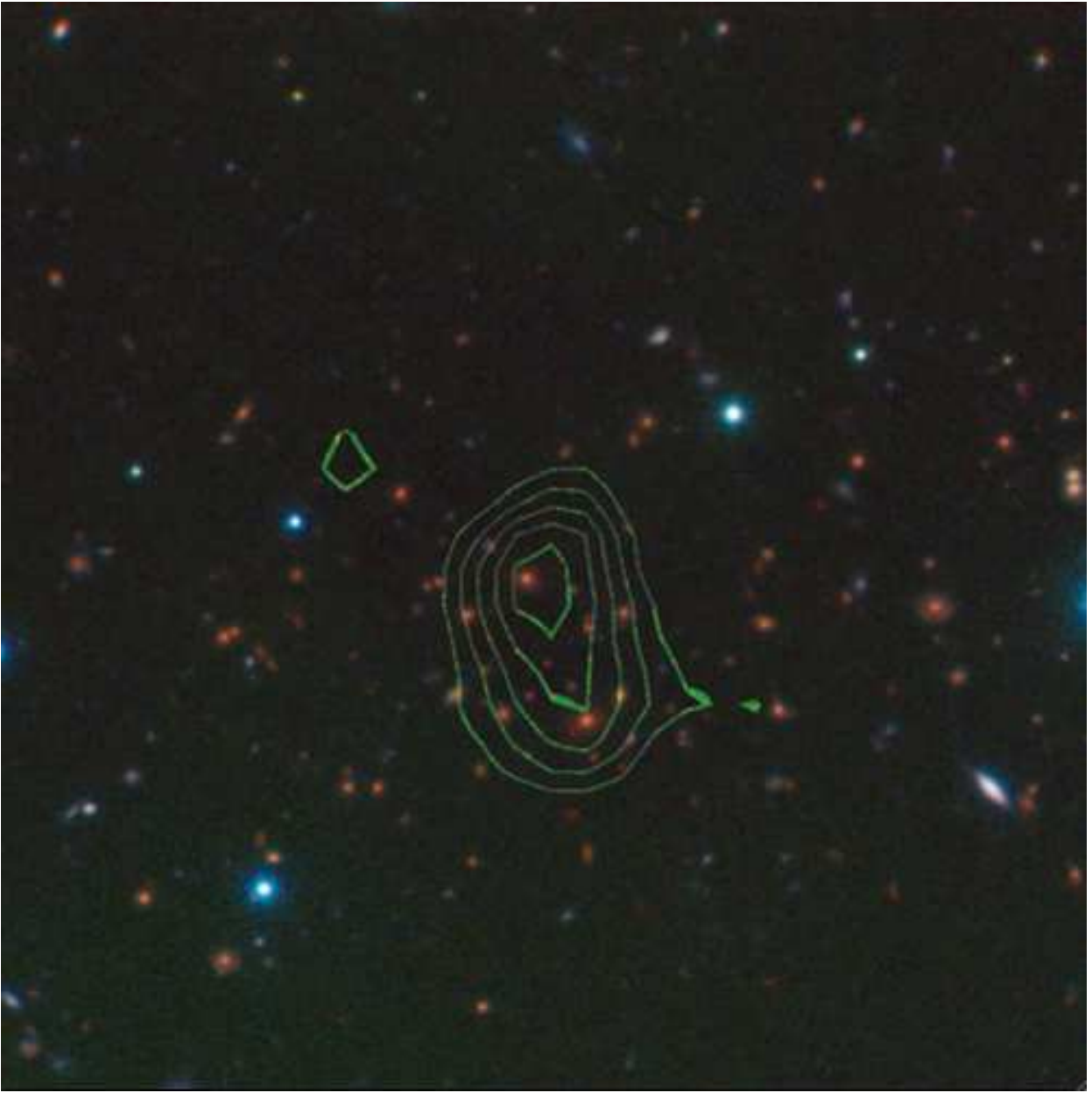}
\caption{Three-color image (Red, green and blue are K, I and R bands, respectively) of \rcsB \ with overlaid green X-ray contours.  North is up and east is to the left, and the image is 2\arcmin\ (797 kpc) on a side.  The X-ray image was smoothed using a 0.75\arcsec\ gaussian kernel and X-ray point sources were removed using the CIAO script \texttt{dmfilth}.  Contours begin $1\sigma$ above the mean counts per pixel (assuming Poisson statistics) and contours are separated by a $0.25\sigma$ spacing.}
\label{rgbfig}
\end{figure}

\begin{figure}[h]
\fbox{
	\plotone{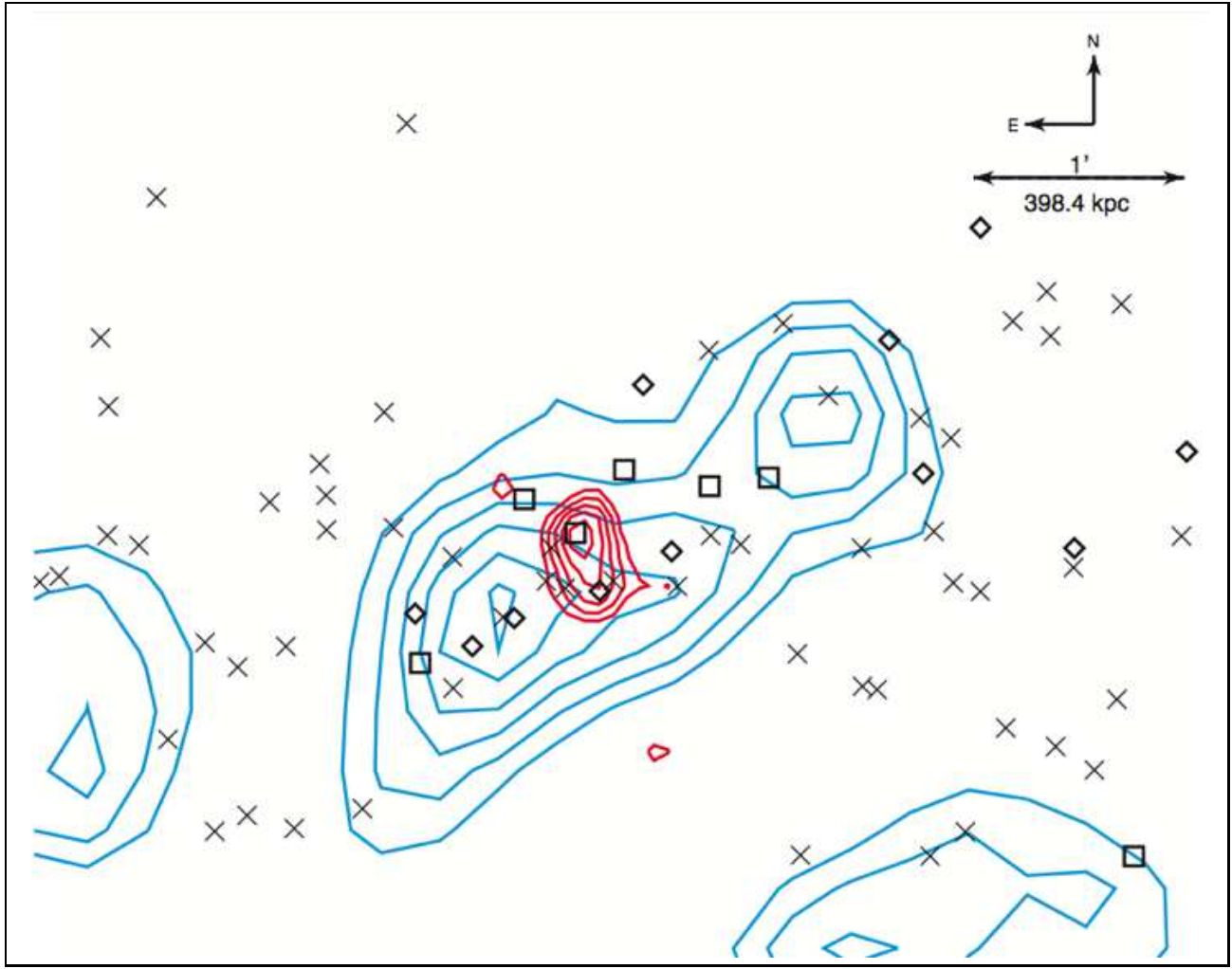}
}
\caption{In red, X-ray contours as in Figure \ref{rgbfig} and in blue RCS galaxy overdensity contours, starting at 2$\sigma$ and increasing in 0.25$\sigma$ increments.  Points indicate galaxies with measured redshifts from optical spectroscopy.  X points indicate galaxies with which are not within a $\pm5000$ \kms\ window of the mean cluster redshift, box points indicate galaxies associated with the $z$=0.94 cluster, and diamond points indicate galaxies associated with the $z$=0.96 cluster.}
\label{contourfig}
\end{figure}

With the new spectroscopy, there are now enough redshifts to attempt to measure approximate velocity dispersions for the two components.  Using all galaxies of redshift quality 4 or better gives: $\sigma=1080\pm320$\kms\ for the $z$=0.9435 component and $\sigma=560\pm160$\kms\ for the $z$=0.9681 component, where the errors are based on jackknife uncertainties as in \citet{Gilbank_ea07}.  Such velocity dispersions would correspond to masses of $7.5\times10^{14}$ M$_{\odot}$ and $1.0\times10^{14}$ M$_{\odot}$ respectively \citep{Cea97}.  However, velocity dispersion estimates based on so few points may be biased and the formal uncertainties are large enough that the values are consistent with both components having equal velocity dispersions.  We note that both peaks in the velocity histogram contain approximately equal numbers of galaxies and use this, coupled with the velocity dispersion uncertainties, to adopt the conservative assumption that both systems have approximately equal masses.  Refined mass estimates from dynamics will require more redshifts and such work is ongoing.  Figure \ref{rgbfig} shows the spatial distribution of these galaxies on the sky in a composite R, I and K band image, overlaid with X-ray contours.  Contours begin $1\sigma$, assuming Poisson statistics, above the average number of counts per pixel in the entire image (which is essentially the background), and are spaced in 0.25$\sigma$ intervals.  Figure \ref{contourfig} shows the same X-ray contours along with RCS galaxy overdensity contours, starting at 2$\sigma$ and spaced in 0.25$\sigma$ increments, as well as the positions of galaxies with measured spectroscopic redshifts, including field galaxies and galaxies from each of the two redshift peaks.  The two redshift peaks overlap in position on the sky and thus it is not possible to say for certain which system is responsible for the X-ray emission measured, or whether both systems contribute, though the latter seems most likely.  From Figures \ref{rgbfig} and \ref{contourfig} it can be seen that the peak of the X-ray emission is nearly coincident with the brightest cluster galaxy (BCG) of the $z$=0.94 cluster, and that the emission is elongated in the direction of the $z$=0.96 cluster's BCG.  This adds further support to the idea that the observed X-ray emission is likely due to a combination of emission from each of the two systems.

Figure \ref{MBfig}, as noted above, shows \bgc\ plotted against M$_{200}$ for our \rcsB\ and a selection of other RCS clusters.   \rcsB\ lies well outside the region bounded by the 1-$\sigma$ scatter, appearing significantly richer than its X-ray mass would suggest.  This is not surprising given that the spectroscopy discussed above suggests that \rcsB\ is an extended structure containing two comparable, less rich systems.  Since these systems lie so close together in redshift that their red sequences overlap, the red-sequence \bgc\ measurement is the sum of that of both these systems.  A more reasonable estimate of the true richness associated with the measured mass X-ray mass, in light of the double-peaked velocity distribution, would be to halve the measured \bgc.  Doing so brings the X-ray mass and the richness mass into agreement, within the large scatter.

\section{Discussion}

\rcsB\ has an inferred X-ray mass which is one to two orders of magnitude lower than the mass indicated by its red-sequence richness, assuming that the X-ray emission is from a single, spherical, isothermal gas distribution and that the richness is associated with a single cluster ($M_{200,X}=4.6\pm^{6.0}_{1.7}\times10^{13}$ M$_{\odot}$ versus $M_{200,B_{gcR}}=1.1\times10^{15}\pm0.46$ dex M$_{\odot}$).  This is a significant discrepancy for two reasons.  First, as mentioned before, richness is well correlated on average with other observable properties.  Secondly, clusters with masses above $10^{15}$ M$_{\odot}$ are quite rare, especially at $z\sim1$.  This means that if the richness mass estimate of \rcsB\ were truly indicative of its size, it would be among the largest overdensities in the observable universe, while also being very underluminous.  Yet the X-ray properties, which suggest a much less massive object, are consistent with the expected $M_{\delta}$--$T_X$ and $L_X$--$T_X$ relations, meaning that the temperature of the plasma is consistent with its observed distribution.  One possible interpretation of these results is that the red-sequence galaxies used to measure the optical richness and the X-ray emitting gas trace different volumes.  Specifically, while the X-ray luminous plasma is confined within one or more deep gravitational potential wells, the galaxy population may occupy a more extended region which is not yet dynamically relaxed.  The existence of such a structure in \rcsB\ is supported by its two-peaked radial velocity distribution.

Merging and dynamically active clusters are expected and observed to be increasingly common at higher redshifts \citep[e.g.,][]{Jea05}.  Thus any cluster sample will include an increasing fraction of dynamically young systems at higher redshift.  Furthermore, N-body simulations indicate that a significant fraction of cluster samples selected using broadband color discrimination will be systems whose galaxy members are not predominantly associated with a single, large potential well, but rather are spread amongst a number of smaller, but still physically associated, dark matter halos distributed along the line-of-sight \citep{Cea07}.  Simulations of the RCS technique on mock catalogs tuned to reproduce observables such as the observed galaxy color distribution and the two-point correlation function show that we expect false-positive cluster detections to occur at a frequency of $\sim$5\% in the RCS \citep{G02}.  This agrees with initial results based on a small number of clusters \citep{Gilbank_ea07,Bea07}.  

Since the red-sequence galaxies used in the RCS richness measurements form very early ($z\gtrsim2$) in high density regions, overdensities of red-sequence galaxies can be associated with very large structures which may not be relaxed during the observed epoch.  The X-ray luminous plasma, on the other hand, will be confined to one or more gravitational potential wells whose size is limited by the collapse timescale.  In the hierarchical collapse paradigm of a $\Lambda$CDM universe, the large, virialized wells seen in local clusters do not develop until long after the galaxies have formed.  In this scenario, for any evolving cluster, there is likely to be an epoch at which the apparent richness overestimates the virialized mass.

Moreover, the misinterpretation of the X-ray emission from a complex, dynamically young cluster as a single, virialized structure can lead to an overestimate of the inferred gas mass fraction and an underestimate of the total mass, which can be seen as follows.  Let us consider two models for the X-ray emission seen in \rcsB.  Our one-component model (Model I) is the set of assumptions used in our X-ray analysis thus far:  X-ray emission is from a single, spherically-symmetric distribution of gas in hydrostatic equilibrium with a gravitational potential which is well-described by a $\beta$-model.  The two-component model (Model II) assumes that there are two nominally identical (i.e.~with the same values of $\beta$, $r_0$ and $T_X$), virialized gas distributions, separated along the line of sight, and that each of the two gas distributions is responsible for half of the measured X-ray flux.  Each component is assumed to obey the local $M_{\delta}$--$T_X$ and $L_X$--$T_X$ relation.  

Because it assumes the flux is halved between the two components, the luminosity inferred per cluster in Model II is half that of Model I.  The luminosity scales as the square of the gas mass, so since the spatial distribution is the same for each cluster, the inferred gas mass and thus also the electron density will be reduced by a factor of $\sqrt{2}$ in each of the clusters relative to Model I.  Since our total mass estimate for each cluster depends only on the temperature of the X-ray emitting gas and its distribution in the plane of the sky, each of the two components in Model II will have the same total mass as the single component in Model I.  This means that the inferred gas mass fraction will be a factor of $\sqrt{2}$ lower using Model II versus Model I.  The total mass, total gas mass and gas mass fraction results using each of the two models can be found in Table \ref{masstable}.

Similarly, $n$ identical components along the line of sight would reduce the inferred gas mass per component and the overall gas mass fraction by a factor of $\sqrt{n}$ in addition to reducing the luminosity per component by a factor of $n$.  In principle, luminosity measurements and the $L_X$--$T_X$ relation constrain the number of components allowed, but our luminosity errors are too large to distinguish between Model I and Model II.

\citet{ME01}, using a suite of hydrodynamic cluster simulations, found that during merger events the ICM spectral fit temperature will underestimate the mass-weighted ICM temperature by $\sim20\%$, because the cool, denser inflowing gas will dominate the emission over the gas already heated by the merger.  This means that for a dynamically young system, such as \rcsB, where even the most virialized components are likely to have undergone recent mergers, X-ray mass estimates for those clusters may still be below the virialized masses by several tens of percent.  In a recent detailed study of the cluster Cl 0024+17, \citet{Jea07} found a similar underestimation of the true mass distribution due to the assumption of a single virialized mass structure rather than two components extended along the line of sight and in a state of ongoing dynamic interaction.  

In summary, the assumption of spherical symmetry and virial equilibrium for a cluster system containing extended line-of-sight structure and dynamic evolution can lead to overestimation of the gas mass fraction and underestimation of the total mass.  The baryon mass fraction in relaxed galaxy clusters is expected to be a universal quantity \citep[e.g.,][]{Wea93,Vea03}, suggesting that the comparison of cluster gas mass fractions inferred by assuming spherical symmetry to canonical cluster values might be used to infer the presence of line-of-sight structure.  As noted by \citet{Wea93}, the cosmological baryon mass ratio, $\Omega_b/\Omega_m$, provides an upper limit to the total gas mass fraction.  This ratio from the WMAP three-year data results is $\Omega_b/\Omega_m=0.175\pm0.012$ \citep{Sea07}.

Galaxy cluster samples at high redshift are expected to have lower cluster gas mass fractions than local samples \citep[e.g.,]{Hea07, Sea05, Eea04}.  Also, a positive correlation between temperature and gas mass fraction has been observed \citep[e.g.,][]{Vea06, Sea03}.  This suggests that a high-redshift cluster with a low X-ray temperature, such as \rcsB, ought to have a correspondingly low gas mass fraction.  Instead, we observe a very high gas mass fraction assuming that it is a single, spherical matter distribution, especially measured within a large radius.  At $r_{500}$,  we find $f_{gas,500}=0.17\pm^{0.07}_{0.08}$.  We observe cluster emission to approximately this radius, so this gas mass fraction estimate is unlikely to have the extrapolation errors that the measurement at $r_{200}$ might.  Assuming that Model II is more appropriate for the physical state of \rcsB\ lowers the gas mass fraction towards what is expected.

Would this pair of clusters be physically associated?  As an order-of-magnitude argument, we note that the richness mass ($\sim$1$\times10^{15}$ M$_{\odot}$) is roughly equal to that of a sphere of radius $\sim$13 Mpc with the cosmic density of $z=0.96$.  Optical spectroscopy shows that the redshift separation of the two components in \rcsB\ is $\Delta z\approx 0.005$, or a physical separation of about 12 Mpc in the Hubble flow.  This rough agreement supports the conclusion that \rcsB\  is an incompletely-virialized system which is still approaching equilibrium, and that the galaxy distribution traces unvirialized matter extended along the line-of-sight, as well as the virialized matter traced by the X-ray gas.  The free-fall collapse time for the two components given by the richness mass is approximately equal to the lookback time, indicating that this cluster would be nearly virialized by about the present epoch.

Additional observations are required to determine the physical state of \rcsB.  Due to the low number of source photons from this cluster, there are large errors in the X-ray luminosity and temperature.  Improvements on both would increase the precision of the X-ray mass estimates and the gas mass fractions.  An independent mass estimate, such as might be obtained from weak lensing, could also help distinguish which mass estimate (richness or X-ray) is more appropriate, as well as giving additional insight into the distribution of matter.  A weak lensing measurement, though difficult, would be particularly interesting for this system if it could trace the spatial distribution of cluster mass perpendicular to the line of sight, possibly revealing more substructure.  Additional spectroscopy, currently underway, will help to further constrain line-of-sight substructure.

\section{Summary}

We present new CXO observations and optical spectroscopy measurements for the high-redshift galaxy cluster \rcsB\ detected in the Red-sequence Cluster Survey.  We have obtained a spectral temperature and estimated total mass, gas mass and gas mass fractions from the X-ray data.  The measured X-ray properties and mass estimates agree well, within errors, to the locally derived $L_X$--$T_X$ and $M_\delta$--$T_X$ relations for galaxy clusters, but not with the local $M_{200}$--$B_{gcR}$ relation.  Spectroscopy of \rcsB\ shows a two-peaked velocity distribution, suggesting substructure along the line-of-sight. Assuming that there is a single, virialized mass distribution in \rcsB\ results in gas mass fraction measurements which are higher than others in the RCS, which may also indicate line-of-sight structure.  Accounting for the substructure with a two-component model for the mass distribution brings the richness mass into agreement with the X-ray mass and the gas mass fractions into agreement with expected values.  Weak lensing measurements, additional X-ray observations and additional spectroscopy would be useful in further constraining both the total amount of mass in the system and its distribution.

\section{Acknowledgments}

BC and MB were supported by NASA via subcontract 2834-MIT-SAO of  contract SV-4-74018
issued by the Chandra X-ray Observatory Center which is operated on  behalf of NASA under 
contract NAS8-03060.  HY acknowledges support from grants from the National Science and Engineering Research Council of Canada and the Canada Research Chair Program.  EE acknowledges support from NSF AST02-06154.  LFB acknowledges the support of the FONDAP center for Astrophysics and CONICYT under proyecto FONDECYT 1040423.



\newpage
\section{Tables and Figures}


\begin{table}[h]
\begin{center}
\begin{tabular}{llll}
\hline
Obsids & Observation Dates$^a$ & Livetime$^b$ &Counts$^c$\\
\hline
\hline
3577, 4438 & 04/16/2003, 06/06/2003 & 93.3/105.0 & 359/2093\\
\hline
\end{tabular}
\end{center}
\caption{Observation Log.  a: Dates are in m/d/y format.  b: Livetime is in ks, and split into good/total amounts.  c: Counts are in the 0.3-7.0 keV band, and split into source/source+background totals.}
\label{obstable}
\end{table}

\begin{table}[h]
\begin{center}
\begin{tabular}{cccccc}
\hline
Redshift & $T_X\ ^a$ & $L_{0.5-2}\ ^b$ & $L_{2-10}\ ^b$ & $L_{bol}\ ^b$ & $S_{0.3-7}\ ^c$ \\
\hline
\hline
0.9558&$1.5 \pm^{1.0}_{0.4}$ &$4.0 \pm^{3.1}_{2.7}$ &$2.1 \pm^{1.7}_{1.4}$ &$6.9 \pm^{4.0}_{3.5}$& $13.7 \pm^{7.5}_{7.0}$ \\
\hline
\end{tabular}
\end{center}
\caption{Spectral Analysis Results.  a: Temperature is in keV.  b: Luminosities in $10^{43}$ erg s$^{-1}$.  Subscripts refer to the rest frame energy band; the bolometric luminosity is estimated by the 0.05-50 keV band.  c:  Measured flux in the 0.3-7.0 keV band, in units of $10^{-15}$ erg cm$^{-2}$ s$^{-1}$.  All values are from within the 49.2\arcsec\ radius spectral aperture.}
\label{spectable}
\end{table}

\begin{table}[h]
\begin{center}
\begin{tabular}{cccccc}
\hline
$I_0\ ^a$ &$n_{e0}\ ^b$  &$r_0\ ^c$ & $\beta$ & $B_{gcR}\ ^d$\\
\hline
\hline
$3.7 \pm^{2.5}_{2.0}$ &$3.9 \pm1.5$ &$14.7 \pm1.9$/$117 \pm15$ &$0.56 \pm^{0.06}_{0.04}$ & 1591$\pm{464}$ \\
\hline
\end{tabular}
\end{center}
\caption{$\beta$-model Results and Optical Richness.  a:  Central surface brightness, in $10^{-14}$ erg cm$^{-2}$ s$^{-1}$ arcmin$^{-2}$.  b: Central electron number density, in $10^{-3}$ cm$^{-3}$.  c: $\beta$-model core radius, in arcseconds/kpc.  d: Optical richness, in $(h_{50}^{-1}$ Mpc$)^{1.77}$, as measured by the RCS \citep{GY05}.}
\label{betatable}
\end{table}

\begin{table}[tbp]
\begin{tabular}{cc|ccc|ccc}
\hline
&&&Model I&&&Model II\\
$\delta$ & $r_\delta\ ^a$ & $M_\delta\ ^b$ & $M_{gas,\delta}\ ^c$ & $f_{gas,\delta}\ ^d$ & $M_\delta\ ^b$ & $M_{gas,\delta}\ ^c$ & $f_{gas,\delta}\ ^d$\\
\hline
\hline
200 & $510\pm^{160}_{70}$ & $4.6\pm^{6.0}_{1.7}$ & $1.05\pm^{0.47}_{0.34}$ & $0.23\pm^{0.09}_{0.11}$ & $9.2\pm^{12.0}_{3.4}$ & $1.48\pm^{0.66}_{0.48}$ & $0.16\pm^{0.07}_{0.08}$ \\
500 & $310\pm^{110}_{50}$ & $2.6\pm^{3.7}_{1.0}$ & $0.45\pm^{0.25}_{0.16}$ & $0.17\pm^{0.07}_{0.08}$ & $5.2\pm^{7.4}_{2.0}$ & $0.64\pm^{0.35}_{0.23}$ & $0.12\pm^{0.05}_{0.06}$ \\
2500 & $90\pm^{65}_{50}$ & $0.33\pm^{1.32}_{0.30}$ & $0.03\pm^{0.08}_{0.03}$ & $0.09\pm^{0.03}_{0.04}$ & $0.7\pm^{2.6}_{0.6}$ & $0.04\pm^{0.11}_{0.04}$ & $0.07\pm^{0.02}_{0.03}$ \\
\hline
\end{tabular}
\caption{Total Mass, Total Gas Mass and Gas Mass Fraction Estimates.  Mass values are for a single component in Model I and for the sum of two components in Model II.  a: The physical radii corresponding to each of the three density contrast radii, in $h_{70}^{-1}$ kpc.  b: Total mass for each model and density contrast, in $10^{13}\ h_{70}^{-1}$ M$_{\odot}$.  c: Gas mass for each model and density contrast, in $10^{13}\ h_{70}^{-5/2}$ M$_{\odot}$.  d: Gas mass fractions for each model and density contrast, in units of $h_{70}^{-3/2}$.}
\label{masstable}
\end{table}


\end{document}